\documentclass[conference]{IEEEtran}
\IEEEoverridecommandlockouts
\usepackage{cite}
\usepackage{amsmath,amssymb,amsfonts}
\usepackage{algorithmic}
\usepackage{algorithm}
\usepackage{graphicx}
\usepackage{textcomp}
\usepackage{xcolor}
\def\BibTeX{{\rm B\kern-.05em{\sc i\kern-.025em b}\kern-.08em
    T\kern-.1667em\lower.7ex\hbox{E}\kern-.125emX}}
\begin{document}

\title{SGCP: A Self-Organized Game-Theoretic Framework For Collaborative Perception
\thanks{
H. Zhang,  Z. Gong, Y. Yang and W. Lu are with the School of Computer Science and Technology, University of Science and Technology of China, Hefei, 230027, P. R. China. (e-mail: {fzhh}@ustc.edu.cn; \{gongzechuan, yuquany, luwenyu9\}@mail.ustc.edu.cn).
}
}

\author{\IEEEauthorblockN{Hui Zhang, Zechuan Gong,  Yuquan Yang and Wenyu Lu}
\IEEEauthorblockA{\textit{School of Computer Science and Technology} \\
\textit{University of Science and Technology of China}\\
Hefei, China \\
{fzhh}@ustc.edu.cn; \{gongzechuan, yuquany,luwenyu9\}@mail.ustc.edu.cn}
}

\maketitle
\begin{abstract}
Collaborative perception holds great promise for improving safety in autonomous driving, particularly in dense traffic where vehicles can share sensory information to overcome individual blind spots and extend awareness. However, deploying such collaboration at scale remains difficult when communication bandwidth is limited and no roadside infrastructure is available. To overcome these limitations, we introduce a fully decentralized framework that enables vehicles to self organize into cooperative groups using only vehicle to vehicle communication. The approach decomposes the problem into two sequential game theoretic stages. In the first stage, vehicles form stable clusters by evaluating mutual sensing complementarity and motion coherence, and each cluster elects a coordinator. In the second stage, the coordinator guides its members to selectively transmit point cloud segments from perceptually salient regions through a non cooperative potential game, enabling efficient local fusion. Global scene understanding is then achieved by exchanging compact detection messages across clusters rather than raw sensor data. We design distributed algorithms for both stages that guarantee monotonic improvement of the system wide potential function. Comprehensive experiments on the CARLA-OpenCDA-NS3 co-simulation platform show that our method reduces communication overhead while delivering higher perception accuracy and wider effective coverage compared to existing baselines.
\end{abstract}

\begin{IEEEkeywords}
Connected and Autonomous Vehicles, Collaborative perception, Vehicular networks, Game theory, Resource allocation
\end{IEEEkeywords}

\section{Introduction}

Connected and Autonomous Vehicles (CAVs) have become a fundamental component of future intelligent transportation systems, empowered by high-resolution sensors such as LiDARs and cameras to support safety-critical functions including collision warning~\cite{lin2015blind} and emergency avoidance~\cite{cui2022coopernaut}. Nevertheless, single-vehicle perception is intrinsically limited by line-of-sight constraints, occlusions and adverse environmental conditions, leading to inevitable perceptual blind spots~\cite{xu2022behind}. These limitations significantly degrade the reliability of autonomous driving, especially in dense urban environments.
collaborative perception (CP) mitigates this issue by allowing multiple CAVs to exchange perception information through V2X communications, thereby extending sensing coverage and improving detection robustness~\cite{wang2020v2vnet,kim2015impact}. In CP, perception tasks are executed collaboratively by a group of CAVs, enabling system-level awareness beyond the capability of any single agent.

However, practical deployment of CP in large-scale vehicular networks is hindered by three fundamental challenges. First, collaborative perception, whether based on raw sensor data or compressed features, incurs substantial communication overhead, which is exacerbated by redundant transmissions from overlapping sensing regions and leads to superlinear growth of channel load as vehicle density increases\cite{liang2025fullperception}. Second, vehicular networks are highly dynamic: mobility, topology and perception quality change rapidly\cite{yang2023what2comm}, rendering static or centralized coordination schemes ineffective under strict real-time constraints. Third, most existing CP systems such as \cite{zhang2021emp,luo2023edgecooper,liang2025fullperception} rely on Roadside Units (RSUs) for centralized scheduling or computing. Such infrastructure is costly, inflexible to traffic dynamics, and often unavailable, making RSU-dependent designs unsuitable for the majority of real-world road networks.

These challenges are exacerbated in dense urban intersections, where tens of vehicles may coexist within a limited area and require perception updates within 100\,ms latency. Meanwhile, current benchmarks such as OPV2V~\cite{xu2022opv2v} and V2XSet~\cite{xu2022v2xset} primarily evaluate sparse scenarios, providing limited insight into the scalability of CP under realistic high-density conditions.

\begin{figure}[htbp]
\centering
\includegraphics[width=1.0\linewidth] {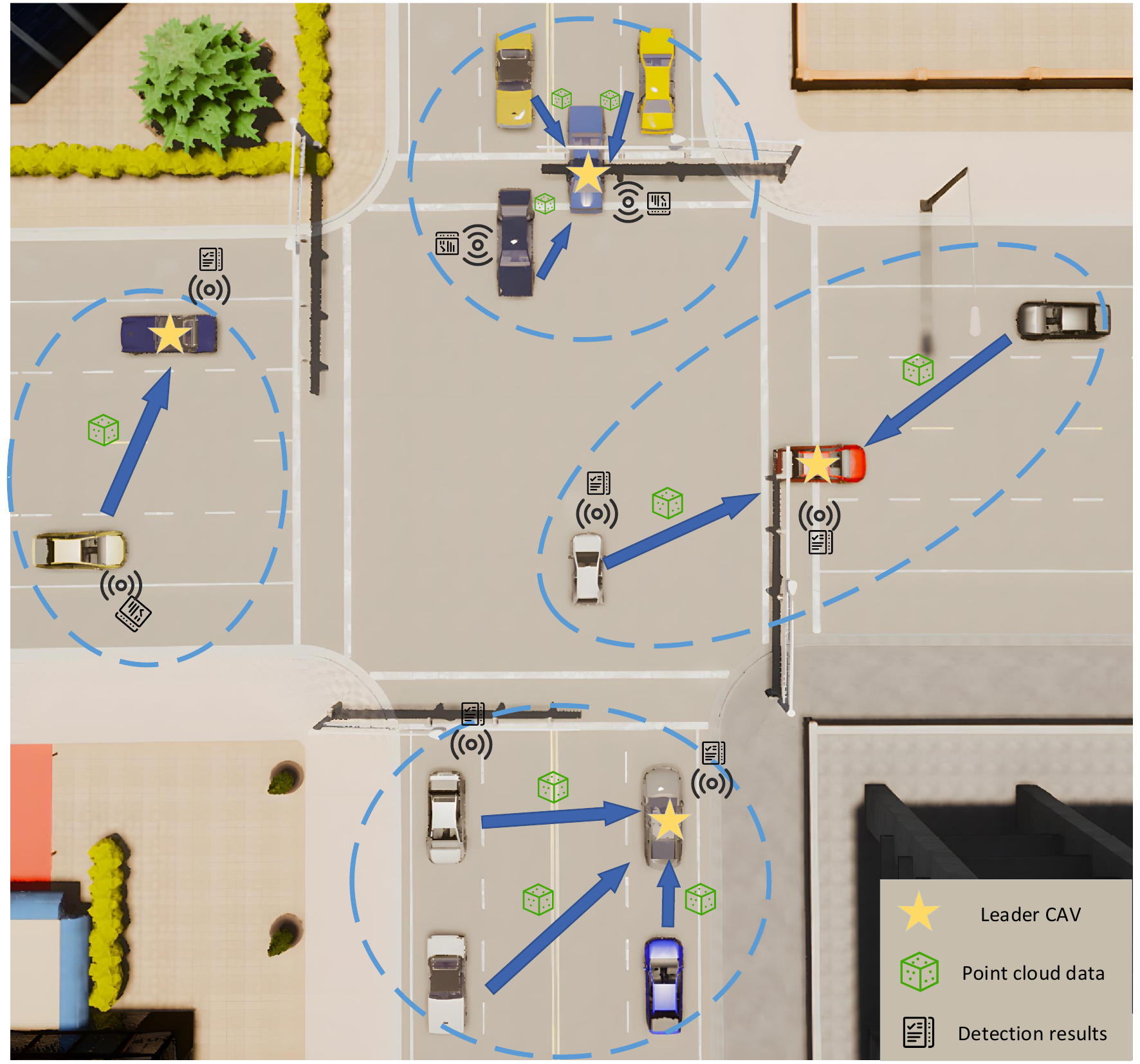}
\caption{System overview: CAVs self-organize into clusters, each with a leader. Members upload high-value point clouds to leaders for early fusion. Cavs broadcast detection results for late fusion across clusters.}
\label{fig:overview}
\end{figure}

To address these limitations, we propose a fully decentralized collaborative perception framework for infrastructure-free vehicular networks, grounded in a self-organized game-theoretic design. Specifically, as depicted in Fig.~\ref{fig:overview}, CAVs autonomously form stable collaboration clusters through a coalition game that evaluates perception performance and motion stability. Within each cluster, the elected leader schedules its members to upload only high-value point cloud regions for early fusion. Across clusters, leaders compete for shared spectrum resources via a non-cooperative potential game, while exchanging lightweight detection results for late fusion. This hierarchical, game-driven approach suppresses redundant data propagation while maintaining high perception fidelity, enabling scalable, infrastructure-free CP in dense urban scenarios.

The main contributions of this paper are summarized as follows:
\begin{itemize}
    \item We propose a fully decentralized cooperative perception framework that eliminates RSU dependency by integrating coalition-game-based cluster formation with potential-game-based resource scheduling, enabling self-organized collaboration under dynamic vehicular conditions.
    
    \item We design a hierarchical fusion architecture that combines intra-cluster point-cloud-level collaboration with inter-cluster result-level collaboration, achieving high perception accuracy while suppressing communication redundancy through perception-aware data selection.
    
    \item We validate the framework on CARLA–OpenCDA–NS3 simulation platform, demonstrating superior performance over state-of-the-art baselines in both perception accuracy and communication efficiency under dense urban scenarios.
\end{itemize}

\section{Related Work}
\label{sec:related}

Collaborative perception (CP) methods can be categorized along two orthogonal dimensions: (1) the granularity of shared data, and (2) the underlying communication and scheduling architecture.

\subsection{Collaboration Granularity}

Based on the level of shared information, CP approaches fall into three categories.

\textbf{Late fusion} exchanges only object-level detection results (e.g., bounding boxes with class and confidence). Arnold et al.~\cite{arnold2020nmslate} applied non-maximum suppression (NMS) to fused detections from multiple vehicles. FusionEye~\cite{liu2019fusioneyelate} integrated multi-view image detections into a unified traffic topology via bipartite matching. Song et al.~\cite{song2023late} proposed a distributed framework using optimal transport for cross-vehicle detection matching, while GraphPS~\cite{li2023graphps} employed graph-based similarity metrics to fuse multi-hop results. late fusion minimizes communication overhead (typically \textless 10\,Mbps~\cite{li2023hetsdn}) and is model-agnostic, enabling seamless cooperation across heterogeneous fleets. However, it cannot recover geometric details lost during individual detection, limiting accuracy in complex scenes requiring precise localization.

\textbf{Intermediate fusion} shares compressed neural network features instead of raw data. V2VNet~\cite{wang2020v2vnet} aggregates features from all neighbors using a graph neural network, improving detection performance but suffering from redundant all-to-all communication. When2Com~\cite{liu2020when2com} introduced a learnable handshake mechanism for dynamic partner selection. Where2Comm~\cite{hu2022where2comm} used spatial confidence maps to guide selective feature sharing, and UMC~\cite{wang2023umc} employed entropy-driven region selection with graph-structured aggregation. Although more efficient than raw data sharing, intermediate fusion requires homogeneous model architectures, hindering deployment in heterogeneous environments~\cite{li2023hetsdn}.

\textbf{Early fusion} directly exchanges raw sensor data, most commonly LiDAR point clouds, and performs detection on the fused global view~\cite{chen2019cooper}. By integrating multi-perspective geometry, it achieves the highest detection accuracy but demands substantial bandwidth (over 100\,Mbps per link~\cite{li2023hetsdn}\cite{liang2025fullperception}) and intensive computation for registration and fusion~\cite{zhang2021emp}. To mitigate these costs, recent works have introduced structured data selection: EdgeCooper~\cite{luo2023edgecooper} employs voxel-based filtering to suppress redundant points, while our work uses grid-based utility-aware uploading.

\subsection{Communication and Scheduling Architectures}

Beyond fusion granularity, the communication architecture critically determines CP scalability and practicality. Nearly all state-of-the-art frameworks rely on Roadside Units (RSUs) for centralized resource management.

EdgeCooper~\cite{luo2023edgecooper} assumes a high-performance RSU that centrally coordinates surrounding CAVs for multi-hop point cloud collection and aggregation. Lu \emph{et al.}~\cite{lu2025joint} allow CAVs to offload partial perception workloads to a single high-capacity RSU and jointly optimize partner selection, data compression, transmission scheduling and computation offloading to balance blind-spot perception quality and end-to-end latency. Zhang \emph{et al.}~\cite{zhao2025multidrl} further treat RSUs and resource-rich vehicles as edge service nodes and leverage deep reinforcement learning for dynamic computation offloading.  FullPerception~\cite{liang2025fullperception} proposes a blind-spot-driven on-demand transmission mechanism, in which the RSU performs global link weight computation, conflict detection and subchannel assignment for centralized resource scheduling, while all perception computation is executed at CAVs.

These RSU-centric designs achieve strong performance in controlled settings but suffer from fundamental limitations: (1) high deployment and maintenance costs make dense urban coverage economically infeasible; (2) fixed RSU locations cannot adapt to spatiotemporal traffic variations; (3) centralized scheduling introduces single points of failure and becomes a bottleneck in high-density scenarios.

In contrast, most existing methods either depend on RSUs or assume idealized communication conditions, limiting their scalability in real-world vehicular networks. To the best of our knowledge, no prior work optimizes collaboration structure and resource-constrained data sharing in a fully decentralized manner under dynamic, large-scale settings. Our framework addresses this gap by integrating stability-aware cluster formation with game-theoretic scheduling to maximize perception gains without roadside infrastructure.

\section{System Model}
\subsection{Decentralized Collaborative Perception Framework}
\label{sec:system_model}
We consider an RSU-free urban intersection scenario,  where a set of CAVs equipped with LiDAR sensors is denoted by $\mathcal{V} = \{v_1, \dots, v_N\}$. 
Each CAV periodically acquires point cloud data at 10\,Hz and participates in collaborative perception. 
Assume that the scene’s $x$–$y$ plane is discretized into a set of non-overlapping grids $\mathcal{G}$ in global coordinates \cite{luo2023edgecooper}\cite{liang2025fullperception}, each CAV $v_i$ observes grids within its sensor radius $R^{\mathrm{sens}}$, forming the sensing region
\begin{equation}
\mathcal{G}_i^{\mathrm{sens}} = \{ g \in \mathcal{G} \mid \|c_g - \mathbf{x}_i\| \leq R^{\mathrm{sens}} \},
\end{equation}
where $\mathbf{x}_i$ is the vehicle position and $c_g$ is the center of grid $g$. 
Similarly, its perceptual requirement region is defined by a larger radius $R^{\mathrm{req}}$:
\begin{equation}
\mathcal{G}_i^{\mathrm{req}} = \{ g \in \mathcal{G} \mid \|c_g - \mathbf{x}_i\| \leq R^{\mathrm{req}} \}.
\end{equation}

CAVs self-organize into disjoint collaboration clusters $\{\mathcal{S}_h\}_{h \in \mathcal{H}}$, where $\mathcal{H} \subseteq \mathcal{V}$ is the set of elected leaders and each cluster $\mathcal{S}_h$ is led by vehicle $h$.  
The collaboration structure defines which vehicles participate in intra-cluster fusion and establishes the scope of collective perception goals. 
For cluster $\mathcal{S}_h$, we define its collective requirement region as the union of all members’ individual requirement regions:
\begin{equation}
\mathcal{G}_{\mathcal{S}_h}^{\mathrm{req}} = \bigcup_{i \in \mathcal{S}_h} \mathcal{G}_i^{\mathrm{req}},
\end{equation}
which represents the total spatial area that the cluster aims to perceive collaboratively.
The decentralized collaborative perception system operates in periodic collaboration cycles of duration $T_c = 100$\,ms, synchronized with the sensing interval. 

\begin{figure}[htbp]
\centering
\includegraphics[width=1.0\linewidth] {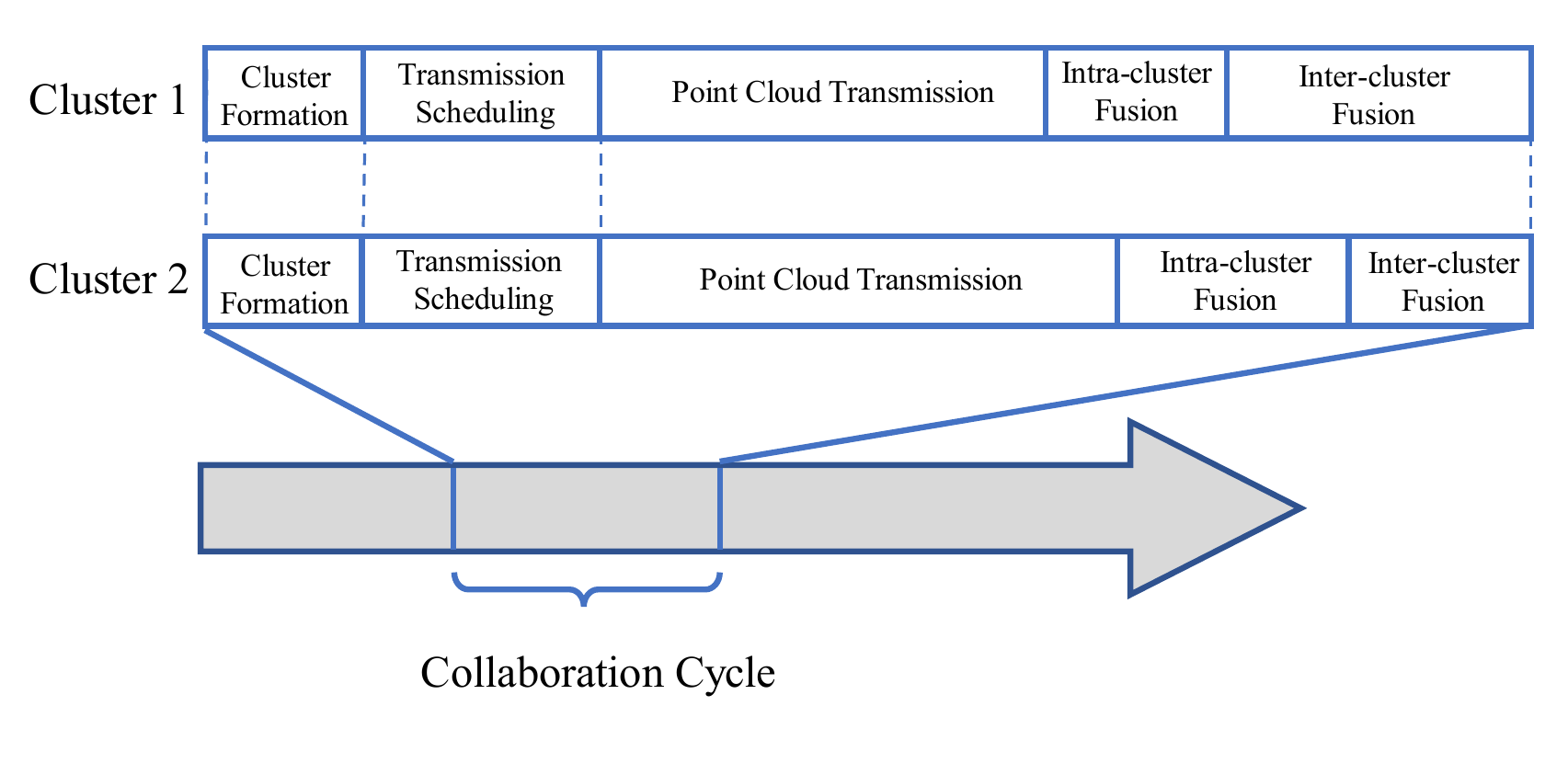}
\caption{Decentralized collaborative perception framework.}
\label{fig:time_slot}
\end{figure}

In the decentralized collaborative perception system, each collaboration cycle contains the following four phases, as shown in Fig. \ref{fig:time_slot}:
\begin{itemize}
    \item \textbf{Cluster Formation}: 
    CAVs self-organize into disjoint clusters based on spatial proximity, motion consistency and sensing overlap. 
    This process is triggered only when significant topology changes are detected (e.g., vehicle entry/exit or large deviation in motion), rather than every cycle. 
    Within each cluster, a leader is elected to serve as the local fusion center.

    \item \textbf{Transmission Scheduling and Point Cloud Transmission}: 
    Each leader determines which grids its members should upload, based on perceptual utility and resource constraints. 
    Only selected grids are transmitted, significantly reducing communication overhead while mitigating intra- and inter-cluster interference.

    \item \textbf{Intra-cluster Fusion}: 
    In each cluster, the leader aggregates received point cloud segments from its members and fuses them into an enhanced multi-view representation. 
    The fused point cloud density over grid $g$ is the sum of contributions from all participating members. 
    This fused density is then processed by a 3D detector (e.g., PointPillars) to produce local perception results.

    \item \textbf{Inter-cluster Fusion}: 
    After local perception is done, cluster leaders broadcast their detection results (object positions, categories and confidence scores) to neighboring vehicles via V2V links. 
    CAVs integrate these results through late fusion to form a consistent global perception.
\end{itemize}

\subsection{Channel Model}
\label{sec:channel_model}
The collaborative perception system employs 5G NR-V2X sidelink communication (PC5 interface) in Mode 2. The available system bandwidth $W$ is divided into $\mathcal K$ orthogonal subchannels, each of bandwidth $B = W / \mathcal K$. 
The instantaneous data rate from CAV $v_i$ to CAV $v_j$ over subchannel $k$ is modeled using the Shannon capacity formula:
\begin{equation}
r_{i,j}^{k} = B \log_2\left(1 + \frac{p_i h_{i,j}^{k}}{\sum_{i' \in \mathcal{V},\, i' \neq i} p_{i'} h_{i',j}^{k} + \sigma^2}\right),
\end{equation}
where $p_i$ denotes the transmit power of $v_i$, $h_{i,j}^{k}$ is the channel gain including path loss, shadowing and small-scale fading, and $\sigma^2$ is the noise power.

Let $l_{i,j}^{k} \in \{0,1\}$ denote the activation status of the V2V link from vehicle $v_i$ to vehicle $v_j$ over subchannel $k$, where $l_{i,j}^{k} = 1$ indicates that the link is active.

The link activation is  subject to half-duplex constraints
\begin{equation}
l_{i,j}^{k} + l_{j,i}^{k} \leq 1, \quad \forall i \neq j,
\label{eq:half_duplex}
\end{equation}
and each active V2V link is assigned at most one subchannel:
\begin{equation}
\sum_{k \in \mathcal{K}} l_{i,j}^{k} \leq 1, \quad \forall i \neq j.
\label{eq:one_subchannel_per_link}
\end{equation}

\section{Problem Formulation}


The quality of collaborative perception is fundamentally shaped by the collaboration structure. To capture this dependency, we first formalize the hierarchical fusion
process under
a given network topology. Building on this model, we then formulate a joint optimization problem
that implicitly determines the optimal cluster formation through scheduling decisions, while
maximizing system-wide perception utility under practical communication and latency constraints.
\subsection{Intra-Cluster Fusion}
\label{sec:intra_cluster}
Within each collaboration cluster, perception quality is determined by the aggregated point cloud density available to the leader. Recent studies confirm that detection accuracy strongly depends on point cloud density: FullPerception~\cite{liang2025fullperception} shows that perception accuracy degrades with sensing distance, which is typically associated with sparser LiDAR point clouds, while EdgeCooper~\cite{luo2023edgecooper} quantifies detection reliability via local point counts. These results establish point cloud density as a fundamental determinant of perception utility.

In sparse regions, insufficient sampling severely degrades classification and localization accuracy~\cite{luo2023edgecooper}. As the sampling density increases, object geometry becomes more complete and detection performance improves rapidly. However, once local structures are sufficiently captured, the performance gain gradually saturates and the marginal benefit of additional points diminishes.

We model detection accuracy as a function $f(\rho)$ of the aggregated point cloud density $\rho$, defined as the number of LiDAR points per unit area (points/m$^2$). The function satisfies
\begin{equation}
f'(\rho) > 0, \quad \lim_{\rho \to \infty} f(\rho) = f_{\max},
\end{equation}
where $f_{\max}$ denotes the upper bound of achievable detection accuracy under a given sensor configuration and scene complexity. Based on this saturating behavior, we define a saturation threshold $\rho_{\mathrm{th}}$ such that $f(\rho_{\mathrm{th}}) \geq (1 - \epsilon) f_{\max}$ for a small tolerance $\epsilon > 0$ (e.g., $\epsilon = 0.05$). Grids with aggregated density $\hat{\rho}_{k,g} \geq \rho_{\mathrm{th}}$ are considered perceptually saturated, and further data transmission to these regions yields negligible utility gain.

The form of $f(\cdot)$ depends on the backbone models and environment and is obtained empirically. In this work, we adopt PointPillars as the backbone model and calibrate $f(\rho)$ using mAP@0.3 measurements from CARLA urban scenarios. This fitted function is treated as system metadata to guide resource allocation, such that communication and computation resources are prioritized for regions with high marginal perception gain, while avoiding waste in saturated areas.

Let $x_{i,j,g}^{k} \in \{0,1\}$ denote the binary decision variable indicating whether CAV $v_i$ uploads the point cloud of grid $g$ to CAV $v_j$ over subchannel $k$. Consistent with the link activation model in Section~\ref{sec:channel_model}, we have
\begin{equation}
l_{i,j}^{k} = \min\Big\{1, \sum_{g \in \mathcal{G}} x_{i,j,g}^{k} \Big\},
\end{equation}

Let $\rho_{i,g}$ denote the raw point cloud density of CAV $v_i$ in grid $g$, with $\rho_{i,g} = 0$ for $g \notin \mathcal{G}_i^{\mathrm{sens}}$. 

The effective point cloud density $\hat{\rho}_{i,g}$ used by vehicle $v_i$ for local detection is then defined as
\begin{equation}
\hat{\rho}_{i,g} =
\begin{cases}
\rho_{i,g} + \sum\limits_{m \in \mathcal{S}_i \setminus \{i\}} \sum\limits_{k \in \mathcal{K}} x_{m,i,g}^{k} \, \rho_{m,g}, & \text{if } v_i \in \mathcal{H}, \\
\rho_{i,g}, & \text{otherwise}.
\end{cases}
\label{eq:rho_hat}
\end{equation}
This formulation indicates that leader vehicles fuse their own observations with scheduled uploads from members, while non-leader vehicles perform local detection using only their own sensor data. The local perception utility derived from grid $g$ is $f(\hat{\rho}_{i,g})$.

The total transmitted data volume from $v_i$ to $v_j$ over subchannel $k$ is
\begin{equation}
s_{i,j}^{k} = \sum_{g \in \mathcal{G}} x_{i,j,g}^{k} \, \rho_{i,g} \, c_0,
\end{equation}
where $c_0$ is the data volume per unit point cloud density.

The corresponding transmission delay is
\begin{equation}
T^{\mathrm{tr}}_{i,j,k} = \frac{s_{i,j}^{k}}{r_{i,j}^{k}}.
\end{equation}

For local computation, let $f_i$ denote the computational capacity of $v_i$ (in FLOPS), and $n_\mathrm{bit}$ the number of floating-point operations per input bit. The computation delay for early fusion at leader $v_i$ is
\begin{equation}
T^{\mathrm{cp}}_{i} = \frac{\sum_{j \in \mathcal{V}} \sum_{k \in \mathcal{K}} s_{j,i}^{k} \cdot n_\mathrm{bit}}{f_i}.
\end{equation}

The perception task must complete within the cooperation cycle $T_c$. For any leader $v_i$:
\begin{equation}
\max_{\substack{j \in \mathcal{S}_i \setminus \{i\} \\ k \in \mathcal{K}}} T^{\mathrm{tr}}_{j,i,k} + T^{\mathrm{cp}}_{i} \leq T_c.
\label{eq:delay}
\end{equation}

\subsection{Inter-Cluster Fusion}
\label{sec:inter_cluster}
Late fusion integrates object-level detection results produced by different CAVs after early fusion and local inference. Each vehicle outputs a set of detected objects, which are subsequently aggregated to form a consistent global perception of the scene.

Due to heterogeneous viewpoints and occlusions, the perceptual quality in a spatial region $g$ varies significantly across vehicles. For a given vehicle $v_k$, the local perception utility in region $g$ is denoted by $f(\hat{\rho}_{k,g})$, where $\hat{\rho}_{k,g}$ is the effective point cloud density used for its local detection.

Late fusion can thus be abstracted as a region-wise aggregation process that preserves the most informative detection result among all contributors. Specifically, let $\mathcal{D}_g \subseteq \mathcal{V}$ denote the set of vehicles that successfully generate and share a detection result for grid $g$. The fused perception utility for grid $g$ is modeled as
\begin{equation}
F_g^{\mathrm{late}} = \max_{k \in \mathcal{D}_g} f(\hat{\rho}_{k,g}),
\label{eq:F_fusion_max}
\end{equation}
which represents the highest detection quality achievable for grid $g$ across the network. Since no detection is possible when the point cloud density is zero, we have $f(0) = 0$. Moreover, for any vehicle $k$ that does not observe or upload data for grid $g$, it holds that $\hat{\rho}_{k,g} = 0$. Consequently,
\begin{equation}
\max_{k \in \mathcal{D}_g} f(\hat{\rho}_{k,g}) = \max_{k \in \mathcal{V}} f(\hat{\rho}_{k,g}).
\end{equation}
This formulation captures two essential properties of late fusion: (i) suboptimal-detection suppression, where low-quality or inconsistent detections are dominated by higher-confidence results, and (ii) coverage enhancement, where observations from any vehicle are propagated to all others through fusion, effectively extending the perceptual field of the system.

Consequently, the total collaborative perception utility of any vehicle $v_i$ is determined by the best detection available for each grid in its requirement region:
\begin{equation}
U_i = \sum_{g \in \mathcal{G}_i^{\mathrm{req}}} F_g^{\mathrm{late}} = \sum_{g \in \mathcal{G}_i^{\mathrm{req}}} \max_{k \in \mathcal{V}} f(\hat{\rho}_{k,g}).
\label{eq:U_total}
\end{equation}
This unified utility model captures the full benefit of hierarchical collaboration: intra-cluster early fusion enhances local detection quality for leaders, while inter-cluster late fusion ensures all vehicles inherit the best available perception across the entire network.

\subsection{Optimization Problem Statement}
\label{sec:problem_formulation}


Based on the perception utility model and the channel constraints, we formulate cooperative perception as a joint optimization problem over cluster formation, communication scheduling and point cloud selection:

\begin{align}
& \max_{\{x_{i,j,g}^{k}\}} \quad \sum_{i \in \mathcal{V}} U_i
\label{eq:target} \\
& \text{subject to:} \quad
\text{
(\ref{eq:half_duplex}), 
(\ref{eq:one_subchannel_per_link})},
(\ref{eq:delay}), 
\notag \\
& \quad x_{i,j,g}^{k} \in \{0,1\}, \quad \forall i,j,g,k. \notag
\end{align}

The objective maximizes the aggregate perception utility across all vehicles, subject to 
end-to-end latency
and physical-layer communication constraints.

This problem is NP-hard. Even the subproblem of selecting non-conflicting V2V links under half-duplex and interference constraints is equivalent to the \emph{maximum weighted independent set} (MWIS) problem on a conflict graph~\cite{basagni2001mwis}. Our formulation further complicates this core scheduling challenge by incorporating (i) nonlinear perception utilities $f(\cdot)$ with saturation effects, (ii) multi-grid point cloud selection decisions. Consequently, exact optimization becomes computationally intractable in large-scale, high-density vehicular networks.

\section{SOLUTIONS}

The joint optimization problem formulated in Section~\ref{sec:problem_formulation} is NP-hard. To enable scalable and distributed coordination in large-scale vehicular networks, we decompose the problem into two subproblems. First, vehicles self-organize into stable collaboration clusters through a coalition game, which determines who should share raw point cloud data for early fusion. Second, within each cluster, leaders perform distributed resource scheduling that jointly selects which point cloud segments to upload and allocates orthogonal subchannels for transmission, as late fusion across clusters operates via direct broadcast of detection results and requires no explicit coordination. This decomposition preserves the hierarchical nature of our framework while ensuring computational tractability.

\subsection{Coalition Game-Based Collaboration Cluster Formation}

Within the proposed hierarchical collaboration framework, self-organized cluster formation in large-scale, dynamic vehicular networks should prioritize coalitions that deliver genuine perception gains beyond what is achievable through late fusion alone. Centralized or proximity-based grouping strategies often fail to distinguish between redundant data sharing and value-adding multi-view fusion. To address this, we model cluster formation as a coalition game in which the value of a coalition $S$ quantifies the marginal benefit of early fusion: specifically, the improvement in perception utility obtained by fusing raw point cloud data within $S$, relative to a baseline where members operate independently and exchange only detection results via late fusion.

\subsubsection{Perception-Based Coalition Value Function}

Formally, for any coalition $S \subseteq \mathcal{V}$, let $\hat{\rho}_{S,g} = \sum_{i \in S} \rho_{i,g}$ denote the aggregated point cloud density over grid $g$, assuming full upload for valuation purposes. The perception utility attainable through early fusion is $f(\hat{\rho}_{S,g})$. In the absence of intra-cluster point cloud sharing, each member performs local detection independently, and the best utility achievable for grid $g$ through late fusion is $\max_{i \in S} f(\rho_{i,g})$, 
according to Section \ref{sec:inter_cluster}.

Consequently, the net gain from forming coalition $S$ is defined as:
\begin{equation}
V(S) = \sum_{g \in \mathcal{G}} \left[ f(\hat{\rho}_{S,g}) - \max_{i \in S} f(\rho_{i,g}) \right].
\label{eq:V_overlap}
\end{equation}

This formulation ensures three desirable properties:
\begin{itemize}
    \item \textbf{Non-negativity}: $V(S) \geq 0$ due to the monotonicity of $f(\cdot)$ and $\hat{\rho}_{S,g} \geq \max_{i \in S} \rho_{i,g}$, guaranteeing that only beneficial coalitions are formed.
    \item \textbf{Multi-view sensitivity}: The gain is maximized when members observe the same grid from geometrically complementary perspectives, rather than merely duplicating similar views, thus encouraging spatially proximate yet angularly diverse clusters.
    \item \textbf{Resource efficiency}: Grids with uniformly low point cloud density across members yield negligible marginal gain, naturally discouraging unnecessary data transmission.
\end{itemize}

Critically, this design aligns with the hierarchical nature of our framework: since late fusion across clusters already provides global coverage and blind-spot compensation, intra-cluster early fusion is reserved exclusively for scenarios where multi-view point cloud fusion yields measurable accuracy improvements beyond late fusion. The resulting $V(S)$ serves as the potential function for our cluster formation algorithm, guiding autonomous vehicles toward clusters that deliver non-redundant, high-value perception enhancements.

\subsubsection{Grid-Based Stability Contribution}

To overcome the short-sightedness of traditional coalition games, we introduce a stability-aware contribution metric that accounts for vehicle motion dynamics.

For coalition $S$, let $\bar{\mathbf{x}}_S$ and $\bar{\mathbf{v}}_S$ denote the average position and velocity of its members. The deviation of CAV $v_i$ is $\Delta\mathbf{x}_i = \mathbf{x}_i - \bar{\mathbf{x}}_S$ and $\Delta\mathbf{v}_i = \mathbf{v}_i - \bar{\mathbf{v}}_S$. Given a minimum stability window $T_{\min}^{\mathrm{stab}}$, its predicted position is
\begin{equation}
\mathbf{p}_i^{\mathrm{pred}} = \mathbf{x}_i + \Delta\mathbf{v}_i \, T_{\min}^{\mathrm{stab}}.
\label{eq:x_pred_vec}
\end{equation}
This prediction is used to estimate the future sensing coverage of $v_i$ and assess its potential overlap with the coalition's requirement region.

The immediate utility gain from $v_i$ joining $S$ is
\begin{equation}
V_i^{\mathrm{early}}(S) = \sum_{g \in \mathcal{G}_i^{\mathrm{sens}} \cap \mathcal{G}_S^{\mathrm{req}}} \left[ f(\hat{\rho}_{S,g} + \rho_{i,g}) - f(\hat{\rho}_{S,g}) \right],
\end{equation}
where $\hat{\rho}_{S,g}$ is the current fused density of coalition $S$.

The future coverage overlap is quantified by the stability coefficient
\begin{equation}
\beta_i = \frac{|\mathcal{G}_i^{\mathrm{pred}} \cap \mathcal{G}_S^{\mathrm{req}}|}{|\mathcal{G}_i^{\mathrm{pred}}|}, \quad 0 \leq \beta_i \leq 1,
\end{equation}
with $\mathcal{G}_i^{\mathrm{pred}} = \{ g \mid \|c_g - \mathbf{p}_i^{\mathrm{pred}}\| \leq R^{\mathrm{sens}} \}$.

The total contribution is then
\begin{equation}
\Delta V_i(S) = \beta_i \cdot V_i^{\mathrm{early}}(S),
\label{eq:deltaV_i}
\end{equation}
which serves as the marginal utility for migration decisions, suppressing frequent reconfigurations while maintaining adaptivity.

\subsubsection{Leader Election and Formation Algorithm}

\paragraph{Leader Election}
Each coalition elects a leader balancing geometric centrality and motion consistency:
\begin{equation}
h^\star = \arg\min_{i \in S} \left( \alpha \|\mathbf{x}_i - \bar{\mathbf{x}}_S\| + (1 - \alpha) \|\mathbf{v}_i - \bar{\mathbf{v}}_S\| \right),
\end{equation}
with $\alpha = 0.7$ prioritizing spatial proximity.

\paragraph{Formation Algorithm}
Coalition size is capped at $N_{\max}$ to avoid congestion. As outlined in Algorithm~\ref{alg:coalition_formation}, each CAV evaluates neighboring coalitions and migrates if $\Delta V_i(S) > 0$. The algorithm maximizes the global potential $\Phi(\mathcal{C}) = \sum_{S \in \mathcal{C}} V(S)$ and converges in finite steps to a Nash-stable partition. Resulting coalitions remain robust over $T_{\min}^{\mathrm{stab}}$, and the procedure repeats every cycle $T_c$ to track dynamics.

\begin{algorithm}[ht]
\caption{Stability-Constrained Cluster Formation and Leader Election}
\label{alg:coalition_formation}
\begin{algorithmic}[1]
\REQUIRE Vehicle set $\mathcal{V}$; positions $\{\mathbf{x}_i\}$, velocities $\{\mathbf{v}_i\}$; sensing grids $\{\mathcal{G}_i^{\mathrm{sens}}\}$; parameters $N_{\max}, T_{\min}^{\mathrm{stab}}$;
\ENSURE Stable coalition partition $\mathcal{C} = \{(S_h, h)\}$
\STATE Initialize: $\mathcal{C} \gets \{ \{i\} \mid i \in \mathcal{V} \}$;
\REPEAT
    \STATE $\textit{updated} \gets \text{false}$;
    \FOR{each vehicle $v_i \in \mathcal{V}$}
        \STATE $S_i \gets$ current coalition of $v_i$;
        \STATE $\mathcal{N}_ i \gets$ neighboring coalitions (e.g., distance $<2R^{\mathrm{sens}}$);
        \STATE $\Delta V_i^{\max} \gets 0$, $S^\star \gets S_i$;
        \FOR{each $S \in \mathcal{N}_i \cup \{S_i\}$}
            \IF{$v_i \notin S$ and $|S| + 1 \leq N_{\max}$}
                \STATE Compute $\Delta V_i(S) = \beta_i \cdot V_i^{\mathrm{early}}(S)$;
                \IF{$\Delta V_i(S) > \Delta V_i^{\max}$}
                    \STATE $\Delta V_i^{\max} \gets \Delta V_i(S)$, $S^\star \gets S$;
                \ENDIF
            \ENDIF
        \ENDFOR
        \IF{$S^\star \neq S_i$}
            \STATE Move $v_i$ from $S_i$ to $S^\star$;
            \STATE $\textit{updated} \gets \text{true}$;
        \ENDIF
    \ENDFOR
\UNTIL{$\textit{updated} = \text{false}$}
\FOR{each coalition $S \in \mathcal{C}$}
    \STATE Elect leader $h^\star$ as above;
    \STATE Record $(S, h^\star)$;
\ENDFOR
\RETURN $\mathcal{C}$;
\end{algorithmic}
\end{algorithm}

\subsection{Distributed Resource Scheduling via Potential Game}

Once the collaboration cluster structure stabilizes, each leader must distributively schedule its members to upload point cloud data over limited time–frequency resources, aiming to maximize its cluster’s perception utility. Since multiple leaders compete for shared wireless spectrum, this problem is naturally modeled as a non-cooperative game. We formalize it as a potential game, where each leader acts as a rational player optimizing its scheduling strategy under communication constraints. The existence of an exact potential function ensures finite-step convergence to a Nash equilibrium, providing a solid foundation for distributed algorithm design.

\subsubsection{Game Modeling and Utility Function}

The players are the set of leaders $\mathcal{H} = \{h_1, \dots, h_H\}$, each representing cluster $\mathcal{S}_h$. The strategy of leader $h$ is its scheduling decision:
\begin{equation}
a_h = \left\{ x_{i,h,g}^{k} \,\middle|\, i \in \mathcal{S}_h,\, g \in \mathcal{G}_{\mathcal{S}_h}^{\mathrm{req}},\, k \in \mathcal{K} \right\},
\end{equation}
where $x_{i,h,g}^{k} \in \{0,1\}$ indicates whether member $v_i$ uploads grid $g$ to $h$ over subchannel $k$, and $\mathcal{G}_{\mathcal{S}_h}^{\mathrm{req}} = \bigcup_{i \in \mathcal{S}_h} \mathcal{G}_i^{\mathrm{req}}$ is the cluster’s collective requirement region.

To ensure consistency with collaboration structure obtained from cluster formation, the strategy space is restricted such that:
\begin{equation}
x_{i,j,g}^{k} = 0, \quad \forall i \notin \mathcal{S}_j,
\label{eq:cluster_constraint}
\end{equation}
\begin{equation}
x_{i,j,g}^{k} = 0, \quad \text{if } v_j \notin \mathcal{H}.
\label{eq:leader_constraint}
\end{equation}
Consequently, for leader $h$, only variables $x_{i,h,g}^{k}$ with $i \in \mathcal{S}_h$ are active; all other entries are fixed to zero by the structural constraints.

According to Section \ref{sec:inter_cluster}, the utility of leader $h$ is defined as the sum of late-fusion utilities over this region:
\begin{equation}
U_h(a_h, a_{-h}) = \sum_{g \in \mathcal{G}_{\mathcal{S}_h}^{\mathrm{req}}} F_g^{\mathrm{late}} = \sum_{g \in \mathcal{G}_{\mathcal{S}_h}^{\mathrm{req}}} \max_{k \in \mathcal{V}}f(\hat{\rho}_{k,g}).
\label{eq:U_total_all}
\end{equation}
This formulation ensures that the leader optimizes resource allocation to maximize the perception quality of the entire cluster.

\subsubsection{Potential Game Properties and Distributed Framework}

The proposed game is an exact potential game. This property arises because each leader’s utility $U_h$ is defined as a sum of terms that depend only on the global maximum perception utility per grid, i.e., $F_g^{\mathrm{late}} = \max_{k \in \mathcal{V}} f(\hat{\rho}_{k,g})$. Since any unilateral change in a leader’s strategy (e.g., uploading additional point clouds) can only increase or leave unchanged the fused density $\hat{\rho}_{h,g}$, and consequently $F_g^{\mathrm{late}}$, the marginal change in $U_h$ exactly matches the marginal change in the system-wide sum of $F_g^{\mathrm{late}}$ over all grids.

Formally, the game admits the following exact potential function:
\begin{equation}
\Phi(a) = \sum_{g \in \mathcal{G}} \max_{k \in \mathcal{V}} f(\hat{\rho}_{k,g}).
\end{equation}
For any player $h$ and any pair of strategies $a_h, a_h'$, it holds that
\begin{equation}
U_h(a_h', a_{-h}) - U_h(a_h, a_{-h}) = \Phi(a_h', a_{-h}) - \Phi(a_h, a_{-h}).
\end{equation}
This equality confirms that $\Phi(a)$ is an exact potential function. As a result, any sequence of best-response updates monotonically increases $\Phi(a)$, and since $\Phi(a)$ is bounded above by $|\mathcal{G}| \cdot f_{\max}$, the dynamics converge in finite steps to a pure-strategy Nash equilibrium.

As shown in Algorithm~\ref{alg:cluster_channel_game}, leaders iteratively update their strategies in parallel by solving local best-response problems based on the current global state. Only lightweight exchange of $\hat{\rho}_{h,g}$ is required.

\begin{algorithm}[ht]
\caption{Perception-Driven Potential Game (PDPG) for Distributed Scheduling}
\label{alg:cluster_channel_game}
\begin{algorithmic}[1]
\REQUIRE Leader set $\mathcal{H}$, cluster structure $\{\mathcal{S}_h\}$, resource pool $\mathcal{K}$;
\ENSURE Converged strategy $a^\star$
\STATE Initialize: Each $h$ generates feasible $a_h^{(0)}$ (e.g., distance-based heuristic);
\STATE $n \gets 0$;
\REPEAT
    \STATE $n \gets n + 1$;
    \FOR{each leader $h \in \mathcal{H}$ \textbf{in parallel}}
        \STATE Observe $\{\hat{\rho}_{k,g}\}_{k \in \mathcal{H}}$, compute $F_g^{\mathrm{late}} = \max_k f(\hat{\rho}_{k,g})$;
        \STATE Solve local best response:
        \[
        a_h^{(n)} \in \arg\max_{a_h \in \mathcal{A}_h} U_h(a_h, a_{-h}^{(n-1)}),
        \]
        where $\mathcal{A}_h$ enforces communication constraints;
    \ENDFOR
\UNTIL{Strategies stabilize or max iterations reached}
\RETURN $a^\star = \{a_h^{(n)}\}_{h \in \mathcal{H}}$;
\end{algorithmic}
\end{algorithm}
\subsubsection{Approximate Best Response: Perception-Priority Scheduling}

The best-response problem is NP-hard, as it involves selecting a subset of members and grids to schedule under a limited resource budget while maximizing a nonlinear perception utility, generalizing the knapsack problem with combinatorial constraints. To enable real-time execution, we propose Perception-Priority Scheduling (PPS), a lightweight greedy approximation that prioritizes perception-critical regions while respecting inter-cluster fairness.

PPS operates under a simple yet effective principle: each cluster is allocated at most $B_h$ subchannels per cycle. The leader first identifies grids where the global perception utility has not yet saturated, i.e., regions where all vehicles exhibit point cloud density below a saturation threshold $\rho_{\mathrm{th}}$. Only these grids are considered for data upload, as additional points in saturated regions yield negligible marginal gain. The leader then computes a perception score for each member, reflecting its potential to improve utility in these under-saturated grids. Members are processed in descending order of this score, and each is assigned one available subchannel until the budget $B_h$ is exhausted.
\begin{algorithm}[ht]
\caption{Perception-Priority Scheduling (PPS): Approximate Best Response for Leader $h$}
\label{alg:best_response}
\begin{algorithmic}[1]
\REQUIRE Cluster $\mathcal{S}_h$, global resource status, $\hat{\rho}_{k,g}$ from all vehicles, saturation threshold $\rho_{\mathrm{th}}$;
\ENSURE Updated scheduling strategy $a_h$
\STATE Initialize: $\mathcal{S}_h^{\mathrm{sched}} \gets \emptyset$, used subchannels $u \gets 0$;
\STATE Construct candidate grid set:
\[
\mathcal{G}_{\mathrm{cand}} = \left\{ g \in \mathcal{G}_{\mathcal{S}_h}^{\mathrm{req}} \,\middle|\, \max_{k \in \mathcal{V}} \hat{\rho}_{k,g} < \rho_{\mathrm{th}} \right\};
\]
\IF{$\mathcal{G}_{\mathrm{cand}} = \emptyset$}
    \RETURN $\emptyset$;
\ENDIF

\FOR{each member $m \in \mathcal{S}_h$ sorted by descending perception score $S_m$}
    \IF{$u \geq B_h$}
        \STATE \textbf{break;}
    \ENDIF
    \STATE Determine $m$’s relevant grids: $\mathcal{G}_m = \mathcal{G}_m^{\mathrm{sens}} \cap \mathcal{G}_{\mathrm{cand}}$;
    \STATE Compute perception score:
    \[
    S_m = \sum_{g \in \mathcal{G}_m} \big[ f(\rho_{m,g} + \hat\rho_{h,g}) - f(\hat\rho_{h,g}) \big];
    \]
    \STATE Assign one idle subchannel to $m$;
    \STATE Schedule all grids in $\mathcal{G}_m$ for upload;
    \STATE Update $\mathcal{S}_h^{\mathrm{sched}}$ and $u \gets u + 1$;
\ENDFOR
\RETURN $\mathcal{S}_h^{\mathrm{sched}}$;
\end{algorithmic}
\end{algorithm}

PPS ensures that each scheduling decision does not decrease the system potential $\Phi(a)$, as it selects only actions with positive marginal utility. The algorithm incurs minimal communication overhead, requiring only the globally shared $F_g^{\mathrm{late}}$ values and local resource occupancy information. Its computational complexity scales linearly with cluster size. By concentrating resources on regions with perceptual bottlenecks, PPS achieves near-optimal system utility with low complexity, making it highly suitable for dynamic vehicular environments.

\section{PERFORMANCE EVALUATION}

\subsection{Experimental Setup and Simulation Platform}

We evaluate the proposed framework on a joint CARLA– OpenCDA–NS3 simulation platform. CARLA~\cite{dosovitskiy2017carla} provides high-fidelity urban intersection scenarios; OpenCDA~\cite{xu2021opencda} integrates the PointPillars perception module\cite{lang2019pointpillars} and implements cooperative logic; NS3~\cite{ns3,ns3-5g-lena} simulates 5G NR V2X Mode 2 D2D links under the 3GPP TR 37.885 channel model.

\begin{figure*}[t]
\centering
\includegraphics[width=0.8\linewidth]{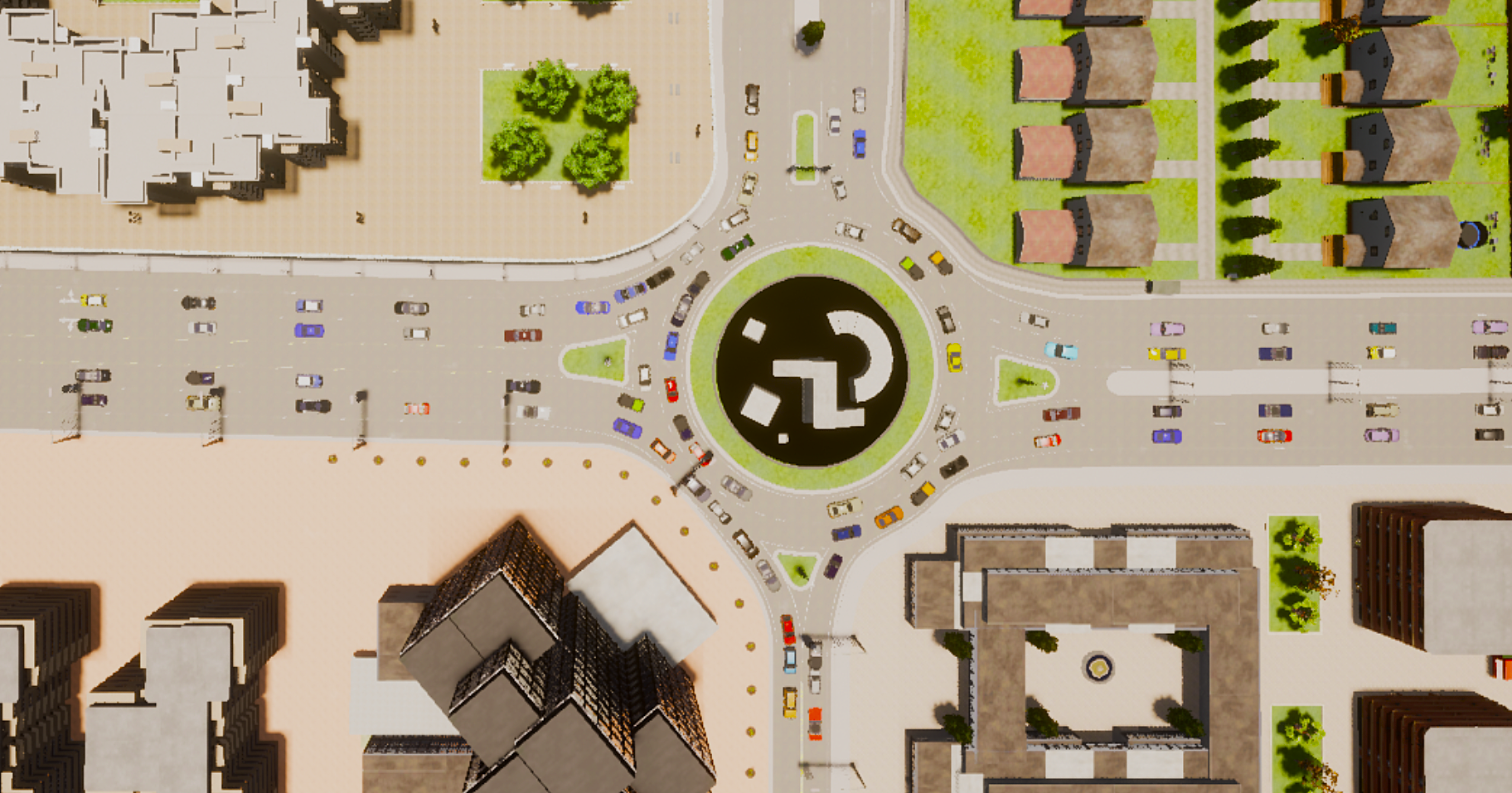}
\caption{A complex urban intersection scenario in CARLA Town03.}
\label{fig:scenario}
\end{figure*}

As illustrated in Fig.~\ref{fig:scenario}, the scenario includes 100 vehicles, of which approximately 20 are CAVs equipped with LiDARs; the rest serve as non-cooperative background traffic. CAVs capture point clouds at 10\,Hz ($\approx$ 56,000 points/s), with a perception range of 50\,m and communication range of 100\,m. Vehicle speeds range from 0 to 60\,km/h. The system operates in the 5.9\,GHz band (n47 TDD) with 40\,MHz bandwidth, divided into $\mathcal{K}=10$ orthogonal subchannels. Each cooperation cycle lasts $T_c = 100$\,ms. Detailed parameters are listed in Table~\ref{tab:sim_params}.

\begin{table}[htbp]
\centering
\caption{Simulation Parameters}
\label{tab:sim_params}
\begin{tabular}{|l|l|}
\hline
\textbf{Channel Parameters} & \textbf{Value} \\
\hline
Frequency band & 5.9 GHz \\
Bandwidth & 40 MHz \\
Number of subchannels ($\mathcal{K}$) & 10 \\
Transmission power & 23 dBm \\
Path loss model & $32.4 + 21 \log_{10}(d_{\text{3D}}) + 20 \log_{10}(f_c)$ \\
Shadowing distribution & Log-normal \\
Shadowing std. dev. & 4 dB \\
Fast fading & Rayleigh \\
Noise power spectral density & $-174$ dBm/Hz \\
\hline
\textbf{Perception Parameters} & \textbf{Value} \\
\hline
Perception range & 50 m \\
LiDAR points per second & 56,000 \\
LiDAR rotation frequency & 10 Hz \\
Communication range & 100 m \\
Number of vehicles & 100 \\
Number of CAVs & 20 \\
Velocity & 0–60 km/h \\
Collaboration cycle & 100 ms \\
Grid size & 10 m \\
Saturation threshold ($\rho_{\mathrm{th}}$) & 2.0 points / m$^2$ \\
Max cluster size ($N_{\max}$) & 4 \\
Stability window ($T_{\min}^{\mathrm{stab}}$) & 500 ms \\
\hline 
\end{tabular}
\end{table}

We determine the saturation threshold $\rho_{\mathrm{th}} = 2.0$ points/m$^2$ through empirical calibration: using PointPillars on a representative CARLA urban intersection scenario, we measure detection accuracy (mAP@0.3) under varying point cloud densities and identify the density beyond which marginal utility gain becomes negligible.

\subsection{Baseline Methods and Evaluation Metrics}

We compare against four baselines:
\begin{itemize}
    \item \textbf{No Cooperation (NC)}: Each CAV runs PointPillars independently;
    \item \textbf{Random Schedule (RS)}: A centralized scheduler randomly selects a conflict-free set of V2V links (satisfying SINR and half-duplex constraints) for point cloud upload;
    \item \textbf{Maximum Utility Greedy (MUG)}: A centralized greedy algorithm that iteratively selects the conflict-free link with the highest marginal perception utility gain;
    \item \textbf{FullPerception}~\cite{liang2025fullperception}: An RSU-assisted cooperative perception framework that adopts blind-spot-driven on-demand transmission and centralized resource scheduling. For fair comparison, we augment it with late fusion.
\end{itemize}
All methods employ PointPillars~\cite{lang2019pointpillars} as the backbone model.

Evaluation metrics include:
\begin{itemize}
    \item \textbf{Perception Performance}: Mean Average Precision (mAP) at different Intersection-over-Union (IoU) thresholds (i.e., mAP@0.3, mAP@0.5 and mAP@0.7);
    \item \textbf{System Communication Overhead}: The aggregate data volume transmitted over all V2V links divided by the experiment duration, measured in Mbps.
    
\end{itemize}

\subsection{Experimental Results Analysis}

\subsubsection{Perception Performance Comparison}

\begin{table}[htbp]
\centering
\caption{Perception Performance (mAP) under Dense Urban Conditions}
\label{tab:mAP}
\begin{tabular}{|l|ccc|}
\hline
Method & mAP@0.3 & mAP@0.5 & mAP@0.7 \\
\hline
NC & 0.13 & 0.12 & 0.10 \\
RS & 0.31 & 0.28 & 0.28 \\
MUG & 0.37 & 0.35 & 0.33 \\
FullPerception & 0.81 & 0.68 & 0.57 \\
\textbf{Ours} & \textbf{0.85} & \textbf{0.84} & \textbf{0.69} \\
\hline
\end{tabular}
\end{table}

Table~\ref{tab:mAP} presents detection accuracy under three IoU thresholds. Our method consistently outperforms NC, RS and MUG across all thresholds. NC achieves only 0.13 mAP@0.3 due to severe occlusions and limited sensing coverage, highlighting the limitations of single-vehicle perception. RS and MUG improve over NC by selecting conflict-free V2V links randomly or greedily; however, they fail to fully exploit communication resources, resulting in constrained detection accuracy and a limited perception range.

Compared to FullPerception, our method achieves relative improvements of 23.5\% at mAP@0.5 (0.84 vs. 0.68) and 21.1\% at mAP@0.7 (0.69 vs. 0.57), demonstrating higher object localization precision. 

These performance gains are attributed to perception-driven self-organized clustering, which prioritizes high-value point cloud regions for intra-cluster early fusion, combined with late fusion that efficiently propagates lightweight detection results across clusters, effectively expanding the perceived area.

\subsubsection{Communication Efficiency Analysis}

\begin{figure}[htbp]
\centering
\includegraphics[width=1.0\linewidth]{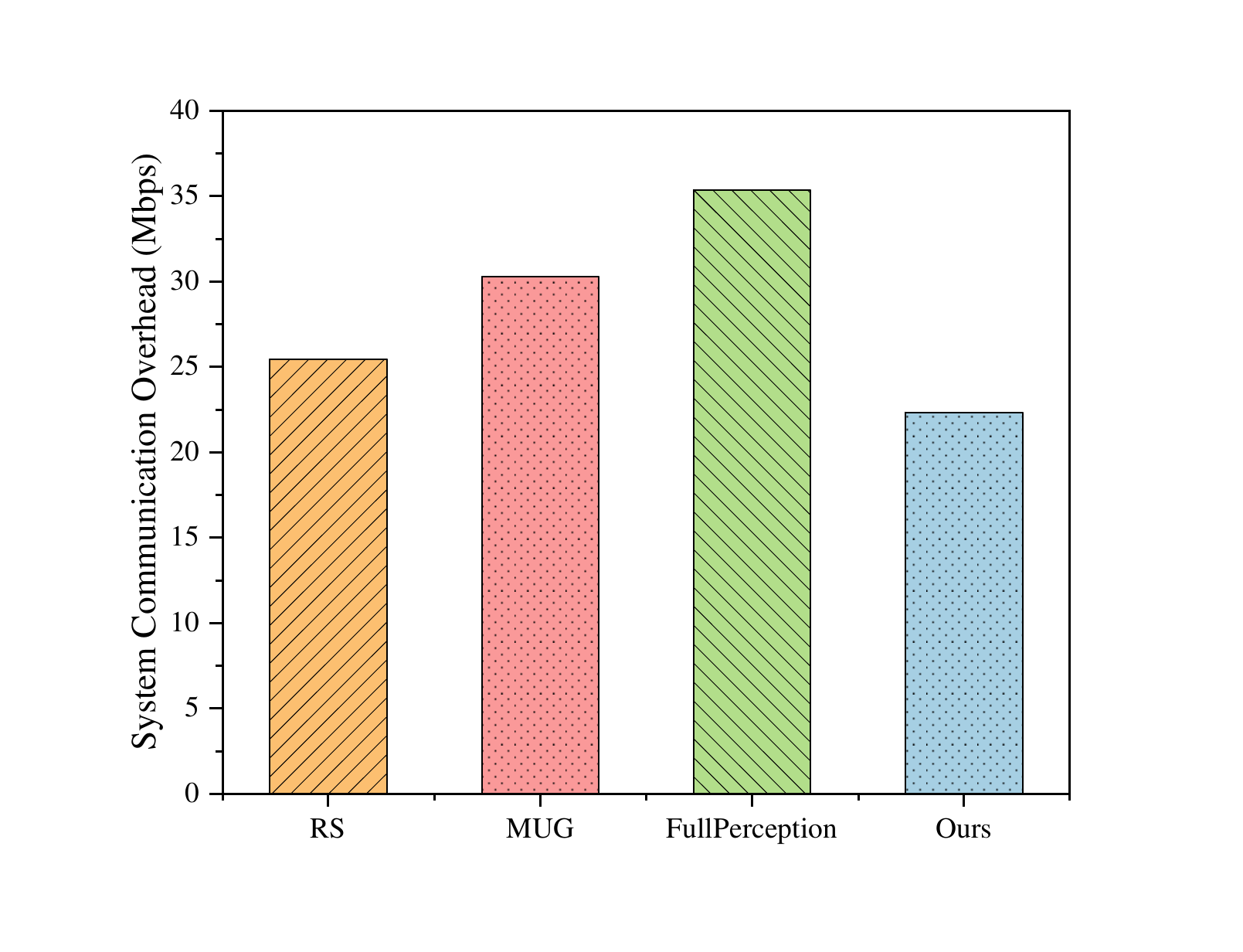}
\caption{System communication overhead (Mbps). Lower values indicate higher communication efficiency.}
\label{fig:comm_overhead}
\end{figure}

Fig.~\ref{fig:comm_overhead} shows system communication overhead in Mbps. Our method requires 22.33 Mbps, lower than FullPerception (35.35 Mbps), MUG (30.27 Mbps) and RS (25.43 Mbps). The improved efficiency arises from our hierarchical fusion design: within clusters, only high-value point cloud regions are uploaded; across clusters, only lightweight detection boxes (\textless 1 KB/object) are exchanged. This combination effectively suppresses redundant transmissions while maintaining high perception accuracy.

\subsubsection{Real-Time Feasibility}
We evaluate runtime performance to ensure the framework meets the 100ms cooperation cycle. Cluster formation converges within 3 iterations per adjustment, and PPS algorithm converges in 3–4 iterations on average. Together with the lightweight late fusion across clusters, the total computation and communication pipeline comfortably fits within the cycle duration.

\section{Conclusion}

This paper addresses the critical challenges of communication inefficiency, RSU dependency and limited scalability in collaborative perception for large-scale vehicular networks. We propose a fully decentralized framework that eliminates infrastructure reliance and achieves high perception accuracy through self-organized V2V collaboration. The core innovation lies in a hierarchical fusion architecture combining intra-cluster early fusion with inter-cluster late fusion, and a two-stage game-theoretic decomposition that jointly optimizes collaboration structure and resource allocation.

Specifically, we formulate the problem as a constrained optimization task and decompose it into two distributed subproblems. First, for collaboration cluster formation, we design a perception-driven coalition game that prioritizes clusters yielding measurable accuracy improvements, augmented with motion-aware stability to suppress frequent reconfigurations. Second, for intra-cluster scheduling, we model resource allocation as a non-cooperative potential game and develop a perception-priority best-response algorithm that selects high-value point cloud regions under physical-layer constraints. Both algorithms are guaranteed to converge to Nash-stable equilibria via monotonic potential functions, and require only lightweight local information exchange.

Extensive evaluations on the CARLA–OpenCDA–NS3 platform demonstrate that our method achieves state-of-the-art performance in dense urban scenarios. Compared to baseline approaches, it improves mAP@0.5 by 23.5\% while reducing communication overhead by 36.8\%. These results validate the framework’s effectiveness in simultaneously enhancing perception fidelity, communication efficiency and system scalability without any roadside infrastructure.

In summary, this work demonstrates that a hierarchical game-theoretic decomposition enables scalable, infrastructure-free collaborative perception in dynamic vehicular networks.

\bibliographystyle{ieeetr}
\bibliography{Reference}

\end{document}